\documentclass[10pt,conference]{IEEEtran}

\usepackage[table, svgnames]{xcolor}
\usepackage[english]{babel}
\usepackage[utf8]{inputenc} 
\usepackage[T1]{fontenc} 

\usepackage[stretch=10]{microtype} 

\usepackage{cite}
\usepackage{amsmath,amssymb,amsfonts}
\usepackage{mathtools}
\usepackage{algorithmic}
\usepackage{graphicx}
\usepackage{textcomp}

\usepackage{caption}
\usepackage{subcaption}
\usepackage{float}

\usepackage{balance}
\usepackage{multirow}

\usepackage{tcolorbox}
\usepackage{listings}
\usepackage{booktabs}
\usepackage[hyphens]{url}
\usepackage[breaklinks=true, colorlinks=true, allcolors=blue, pdfencoding=auto, psdextra, pagebackref, linktoc=all]{hyperref}
\usepackage{bookmark}
\ifdefined\backref
\renewcommand*{\backref}[1]{}
\renewcommand*{\backrefalt}[4]{%
	\ifcase #1 (Not cited.)
	\or  (Cited on page:~#2)
	\else  (Cited on pages:~#2)
	\fi%
}
\else
\fi

\usepackage{siunitx} 
\usepackage[autostyle=true,english=american]{csquotes} 
\MakeOuterQuote{"}

\def\BibTeX{{\rm B\kern-.05em{\sc i\kern-.025em b}\kern-.08em
 T\kern-.1667em\lower.7ex\hbox{E}\kern-.125emX}}
\usepackage{algorithm}
\usepackage{algorithmic}
\usepackage{todonotes} 
\usepackage{cleveref} 
\crefname{lstlisting}{listing}{listings} 
\Crefname{lstlisting}{Listing}{Listings} 
\Crefname{figure}{Fig.}{Fig.}
\Crefname{equation}{Eq.}{Eq.}

\usepackage{soul}
\sethlcolor{WhiteSmoke}

\begin{document}

\makeatletter
\newcommand{\linebreakand}{%
  \end{@IEEEauthorhalign}
  \hfill\mbox{}\par
  \mbox{}\hfill\begin{@IEEEauthorhalign}
}
\makeatother

\title{Fault Localization via Fine-tuning Large Language Models with Mutation Generated Stack Traces}

\author{
\IEEEauthorblockN{Neetha Jambigi}
\IEEEauthorblockA{University of Cologne \\
Cologne, Germany }
\IEEEauthorblockA{njambigi@smail.uni-koeln.de}
\and

\IEEEauthorblockN{Bartosz Bogacz}
\IEEEauthorblockA{ 
SAP\\ Walldorf, Germany\\ bartosz.bogacz@sap.com
}
\and
\IEEEauthorblockN{Moritz Mueller}
\IEEEauthorblockA{ 
SAP\\ Walldorf, Germany\\moritz.mueller07@sap.com
}
\and

\IEEEauthorblockN{Thomas Bach}
\IEEEauthorblockA{SAP \\ Walldorf, Germany\\
0000-0002-9993-2814}

\linebreakand
   
\IEEEauthorblockN{Michael Felderer}
\IEEEauthorblockA{German Aerospace Center \\ University of Cologne \\
Cologne, Germany \\
0000-0003-3818-4442}

}

\maketitle

\begin{abstract}

Abrupt and unexpected terminations of software are termed as software crashes. 
They can be challenging to analyze. Finding the root cause requires extensive manual effort and expertise to connect information sources like stack traces, source code, and logs.
Typical approaches to fault localization require either test failures or source code. Crashes occurring in production environments, such as that of SAP HANA, provide solely crash logs and stack traces. 
We present a novel approach to localize faults based only on the stack trace information and no additional runtime information, by fine-tuning large language models (LLMs). We address complex cases where the root cause of a crash differs from the technical cause, and is not located in the innermost frame of the stack trace. As the number of historic crashes is insufficient to fine-tune LLMs, we augment our dataset by leveraging code mutators to inject synthetic crashes into the code base. 
By fine-tuning on 64,369 crashes resulting from 4.1 million mutations of the HANA code base, we can correctly predict the root cause location of a crash with an accuracy of 66.9\% while baselines only achieve 12.6\% and 10.6\%.
We substantiate the generalizability of our approach by evaluating on two additional open-source databases, SQLite and DuckDB, achieving accuracies of 63\% and 74\%, respectively. Across all our experiments, fine-tuning consistently outperformed prompting non-finetuned LLMs for localizing faults in our datasets.

\end{abstract}




\section{Introduction}
\label{s:introduction}
Fault localization in software engineering entails identifying parts of code that lead to failures~\cite{wong2016survey}. It is a crucial step in software debugging. The effort required to manually locate faults may increase proportionally to the size and complexity of the software projects. Automatically locating the part of the code containing the problem with reasonable effectiveness can significantly accelerate issue resolution.

In the case of SAP HANA, a database management system
for enterprise applications developed by SAP~\cite{farber2012sap,farber2012sap1}, the
source code consists of about 40 million lines of code and
is changed about 200 times a day, making it impossible for
a single person to understand all parts and all modifications~\cite{bach2022testing}. The tests in continuous integration infrastructure mostly aim at identifying bugs and possible regressions. However, software crashes occurring in production scenarios violate most of the assumptions in unit and functional tests. The crash, in most cases, is reported in the form of a bug report augmented with logs and stack traces resulting from a crash. In such a scenario, there are no failing test cases or source code elements accessible for automated fault analysis or localization.  Even with the availability of source code, it still takes a lot of effort to narrow down exactly the faulty code element. Therefore, an automatic fault localization mechanism to identify the faulty code elements is beneficial.

Stack traces have continually proven to be helpful for fault localization~\cite{wong2014boosting, sinha2009fault, kang2023preliminary}.  Previous works have focused on empirically assessing the importance of stack traces in debugging, fault localization~\cite{wong2014boosting, sinha2009fault, kang2023preliminary, gong2014locating} and their effect on bug resolution times~\cite{schroter2010stack}.  Stack traces are extensively utilized in addition to other factors like test cases, program code, etc., in automated fault localization~\cite{zou2019empirical}.  Although manually analyzing stack traces is valuable for diagnosing problems it can be time-consuming. Automating this process improves efficiency by quickly identifying patterns and reducing debugging time and effort. In the bug benchmark Defects4J~\cite{just2014defects4j}, only 3.33\% of the crashes contain crash-triggering test cases~\cite{barreto2024leveraging}, which further emphasizes an approach to localize faults based on stack traces.


Fault localization has conventionally been done using criteria,  such as failing test cases and their coverage information ~\cite{jones2005empirical, wong2013dstar} or by applying specific mutation operators to the original code elements and analyzing the mutants' execution. However, the capabilities of large language models (LLMs) have been increasingly
leveraged in the field of software engineering for various tasks, including but not 
limited to code summarization~\cite{ahmed2022few}, code generation~\cite{yetistiren2022assessing}, test case generation~\cite{guilherme2023initial}, automatic program
repair~\cite{joshi2023repair}, or fault localization~\cite{yang2024large, kang2023preliminary}.
Recent approaches to fault localization using LLMs
regularly outperform conventional statistical and deep learning-based methods~\cite{kang2023preliminary, wu2023large}. Unlike classical machine-learning methods, LLMs excel in multi-modal learning with free-form text~\cite{zhao2023survey}, including program text, stack traces, and SQL queries. They can be adapted to domain-specific tasks using one-shot or few-shot prompting, fine-tuning, and instruction tuning~\cite{zhao2023survey}. Effective LLM fine-tuning needs large, task-specific datasets. While authentic crashes from real-world systems are the gold standard to learn from, they are rare with very limited samples per code element, which is not enough to fine-tune LLMs. They may not cover all code elements. Therefore, for fault localization in crashes, mutation-based techniques can generate relevant training data in the form of stack traces of crashes.

 In this work, we utilize pre-trained open-source LLMs to localize faults using stack traces. We generate a new dataset by mutating the source code of large database projects and triggering crashes during test suite execution. We make the following contributions:
\begin{itemize}

\item A novel approach to localize faults by solely utilizing stack traces.
 \item An evaluation with different LLMs across multiple codebases that shows the viability and generalizability of our approach.
\end{itemize}


The paper is structured as follows: \Cref{s: related_work} presents related work. \Cref{s: approach} describes our approach using stack traces and fine-tuning open-source LLMs for fault localization. \Cref{s: data} outlines our mutation-based process to generate failure data for fine-tuning. \Cref{s: prediction_task} presents prediction methods with both fine-tuned and non-fine-tuned LLMs. \Cref{s: expt_and_eval} details experiments on HANA, SQLite, and DuckDB crashes.  \Cref{s: discussion} discusses results on synthetic and authentic crashes and the approach’s limitations. \Cref{s: conclusion} concludes and presents future directions for our work.

\section{Related Work}
\label{s: related_work}






\subsection{Statistical and Mutation Based Methods}
Statistical methods like Spectrum-based fault localization~\cite{de2016spectrum} utilize coverage metrics from failing test cases to assign suspiciousness scores to the elements in code. There are various methods~\cite{jones2005empirical, wong2013dstar} to calculate these metrics. Localizing crashes by leveraging stack traces leads to better efficacy of spectrum-based methods ~\cite{barreto2024leveraging}. Mutation-based fault localization methods apply mutations to the program, and the failures from test cases that cover the code path of the mutations are used to localize the faults~\cite{moon2014ask, papadakis2015metallaxis}. Both these methods are impacted by the test suite's comprehensiveness and its ability to detect a wide range of faults. Furthermore, these methods are contingent upon failing test cases, where flakiness can play a role~\cite{sarhan2022survey}. False positives from these methods can complicate fault localization efforts further.



\subsection{Deep Learning and LLM Based Methods}

Learning-to-rank faulty code elements involves scores from other fault localization methods, including spectrum-based methods~\cite{xuan2014learning}, mutation-based methods, or a hybrid of both~\cite{li2017transforming}. Deep learning-based approaches may further combine multimodal factors like code complexities, test results, and code structures.  DeepFL~\cite{li2019deepfl} combines scores from the spectrum and mutation-based methods and textual attributes of the code to localize faulty code elements. GRACE~\cite{lou2021boosting} encodes the coverage information through graph representation to rank the program elements. FixLocator~\cite{li2022fault} localizes co-changing faults across the function and statement level using a Graph Convolutional Network encoding program dependencies. DEEPRL4FL~\cite{li2021fault} find representation for coverage matrix and TRANSFER-FL~\cite{meng2022improving} combines spectrum, mutation-based score, and semantic features from code to localize faults at the statement level. These approaches outperform conventional methods but need substantial data to train the models on. In contrast, pre-trained LLMs possess these abilities out-of-the-box and have better generalizability.


Pre-trained LLMs perform a wide range of software engineering tasks more effectively than traditional methods~\cite{hou2023large, fan2023large}. LLM-based fault localization in AutoFL~\cite{kang2023preliminary} outperforms both Spectrum-based~\cite{jones2005empirical, wong2013dstar} and Mutation based~\cite{moon2014ask, papadakis2015metallaxis} methods by iteratively prompting the GPT-4~\cite{OpenAI_GPT4_2023} with a combination of failing test case code, stack traces, and code coverage to localize faulty lines. LLMAO~\cite{yang2024large} extends an open-source LLM and exploits the naturalness of code for localizing faults based only on source code.  Automatic program repair approaches perform fault localization as an intermediate step and leverage LLMs to generate repair. The Inferfix~\cite{jin2023inferfix} framework proposes program repair utilizing an LLM and localizes fault using static analyzer information. RING~\cite{joshi2023repair} considers a suggestion of repair for an appropriate location as fault localization. OpenAI's closed-source models~\cite{bubeck2023sparks, chen2021evaluating} limit exploration to prompting strategies, and can be prohibitively expensive~\cite{deldjoo2023fairness}.  

For proprietary software like HANA, there is insufficient usable failure data for these models to perform fault localization effectively. Most of the existing methods focus on scenarios where more than just the crash information is available and cannot leverage LLMs, especially for closed-source software systems like SAP HANA. Therefore, evaluating fine-tuning open-source models for fault localization is valuable.

\section{Approach}
\label{s: approach}

In this section, we provide an overview of our approach while motivating our choices. Essentially, our approach involves using an LLM fine-tuned with stack traces to predict the name of the function causing the crash. Additionally, this fine-tuned language model, when used in production environments, can learn to map a new crash that partially or fully resembles known instances, and then use this information to localize crashes by interpolating mappings to the nearest code elements.

We present a framework for using mutation-induced crash data to fine-tune LLMs for automated fault localization in software crashes. Authentic crashes that can be collected from real systems are a rather small set. To address the scarcity of real crash data, we propose generating diverse crash scenarios through code mutations based on test coverage. This dataset is then used to fine-tune an LLM, enhancing its ability to identify the code elements responsible for new crashes. Our approach compensates for data limitations and leverages the strengths of LLMs to improve automated fault localization.



\subsection{Mutation of Source Code}
Mutation testing relies on mutations to detect real faults and evaluate the robustness of the test suites. Researchers have already examined the effectiveness of mutations in identifying real faults in both developer-written and auto-generated test cases~\cite{just2014mutants}.
Additionally, simple mutations are capable of identifying complex failures~\cite{gay2023closely}. Based on these findings, we adopt mutations as a means of inducing program crashes to collect failure data. 
When programs are mutated, they execute code in unexpected ways, often violating internal assumptions and following unconventional code paths.   An example of a mutation changing the condition in the ternary operator is shown in \Cref{fig: promptExample}. The resulting stack traces from mutated code can be considered as an incomplete record of the abnormal program flow, showing only the current path in the execution of the program. If mutated program execution is framed as a function that maps code mutations to eventual crashes, our approach can be understood as a sparse sampling of this mapping and using an LLM as an interpolator to fill in the inverse i.e., crashes to mutations.

\subsection{Stack Traces as Failure Information}
Crashes caused by mutations of the source code or otherwise typically result in a stack trace. Stack traces are regarded as one of the important pieces of information available to developers to navigate the code to locate errors~\cite{schroter2010stack}. They contain a list of active stack frames during program execution. Each frame represents a function call that is still in progress. The stack trace may or may not contain the method name which may have been the root cause of the failure. The presence or absence of the root causes in the stack traces are symptoms of crash stacks being incomplete descriptions of program execution.  They are only a snapshot of the call sequence leading to a crash but may miss the broader sequence of events and interactions that contribute to the issue. Despite this fact, we posit that we can leverage such incomplete information and localize crashes that occur long after an incorrect state is introduced in program execution. Such crashes are typically hard to localize manually.

\subsection{Fine-tuning LLM}
Pretrained LLMs encapsulate knowledge from their training data. Decoder-only LLMs~\cite{chen2021evaluating, fan2023large} consist solely of the decoder part of the transformer architecture and are used for sequence generation tasks in natural language processing. They are adept at completing documents by generating tokens based on contextual information. Fine-tuning adapts the LLMs to domain-specific tasks by further training the LLMs on data mapping useful for the domain. In our case, the context is stack traces and as a completion task, the LLM is expected to generate the function name.

\section{Data}
\label{s: data}

  We synthesize artificial data to fine-tune the LLMs to perform fault localization. The data creation process consists of mutating the source code and utilizing the test suites to execute the mutated source code. 

\begin{figure*}[htbp]
\centering
    \includegraphics[scale=0.95]{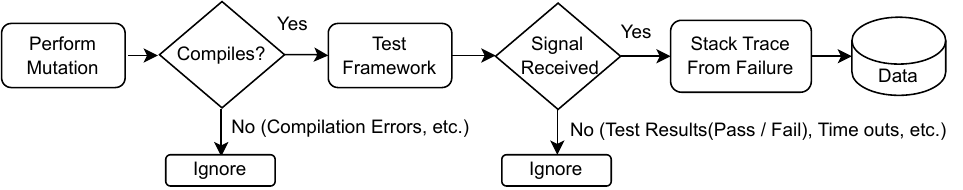}
\caption{Data creation process.}
\label{fig:MutationProcess}
\end{figure*}


 Fine-tuning requires a large dataset with clear correspondence between code changes and their resulting stack traces. Defects4J~\cite{just2014defects4j} and BugsInPy~\cite{widyasari2020bugsinpy} are popular open-source bug benchmarks containing bugs with fixes in singular files. The problem with using these datasets is that most LLMs are trained on these public datasets, incorporating knowledge from these benchmarks. Their sizes are too
limited for fine-tuning. These datasets contain multiple codebases and provide a maximum of 174 bugs per codebase. These benchmarks have bugs that span diverse projects over extended lifetimes containing multiple build configuration changes. Therefore, automating mutation with bug benchmarks has proven to be highly challenging in our initial experiments. Instead to evaluate our approach, in addition to SAP HANA, we mutate two well-known open-source projects, SQLite and DuckDB, which offer quick build and test turnaround times.  We synthesize a labeled dataset by artificially injecting faults into a codebase (i.e., mutating it) so that a clear correspondence between fault location and crash stack trace can be established and learned by the LLM.

\subsection{Synthetic Data Creation}
We perform coverage-guided mutations to trigger potential crashes and rely on test suites to execute mutated code as shown in \Cref{fig:MutationProcess}.   A coverage-guided mutation is effective only on code that is reachable by existing test cases. However, a limitation of this approach is that faults can exist in both tested and untested code paths, potentially leaving some crashes excluded from our dataset.
Generally, more actively used code paths and those with a history of crashes tend to have better test coverage. Regardless, a dataset of crashes generated through source code mutations will generally present faults from a more varied range of code paths compared to datasets based solely on real-world crashes.



 




\subsection{Mutators}

\begin{table}[]
\caption{Example of mutators.}
\resizebox{\columnwidth}{!}{
\begin{tabular}{@{}lll@{}}
\toprule
Name               & Operation                    & Set                                                                                                                                                              \\ \midrule
Assignment         & Replace with symbol from set & = += -= *= /= \%=                                                                                                                                                \\
Number             & Increment by value           & +255 +1 -1 -255 *(-1)                                                                                                                                            \\
LineOrder          & Swap line with random other  &                                                                                                                                                                  \\
BooleanAssignment  & Replace with symbol from set & = \&= |= \textasciicircum{}=                                                                                                                    \\
Delete             & Blank line                   &                                                                                                                                                                  \\
Comparison         & Replace with symbol from set & == != \textless{}\textgreater{} \textbackslash{}\textless{}= \textbackslash{}\textgreater{}=                                \\
Symbol             & Replace random other symbol  &                                                                                                                                                                  \\
Arithmetic         & Replace with symbol from set & + - * / \%                                                                                                                                                       \\
IncrementDecrement & Replace with symbol from set & ++ --                                                                                                                                                            \\
BooleanArithmetic  & Replace with symbol from set & \& | \textasciicircum{} \textless{}\textless{} \textbackslash{}\textgreater{}\textgreater{} \\
Logical            & Replace with symbol from set & \&\& and || or != not                                                                                                                                            \\ \bottomrule
\end{tabular}
}

\label{tab: Mutators}
\end{table}

\begin{figure*}[ht]
        \centering
        \includegraphics[scale=0.9]{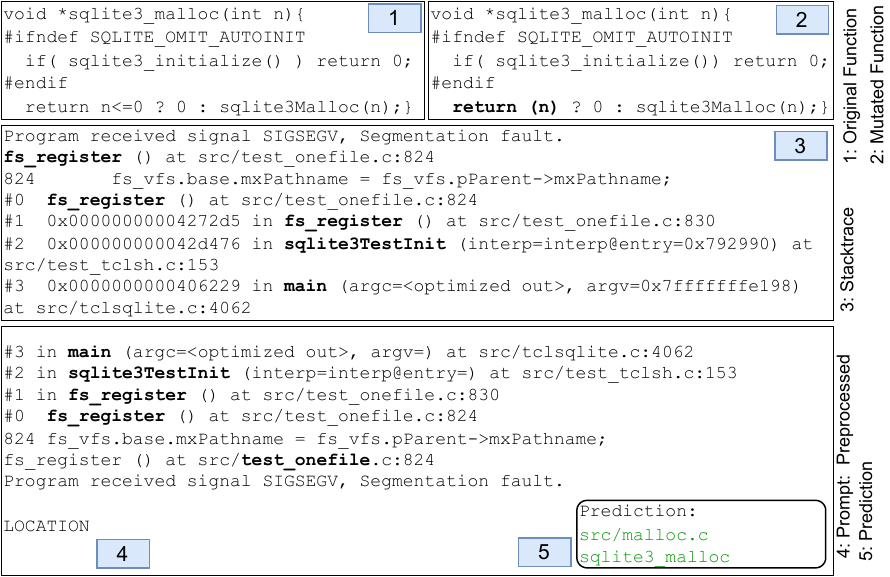} 
        \captionof{figure}{Example of Non-local Crash due to Mutation to \textit{malloc} function in SQLite. This figure outlines mutation, pre-processing stack traces, fine-tuning prompt, label and prediction steps in our pipeline.}
        \label{fig: promptExample}
\end{figure*}



We apply mutations to create variants of the program that are almost identical to the original but contain one minor defect.  These mutators cause code-path divergences, resulting in errors of varying complexity. When the mutated method is present in the stack trace of the crash, it can be easier to locate such faults. However, \Cref{fig: promptExample} shows a simple mutation causing a complex failure that crashes outside the mutated method, which does not appear in the stack trace. Such failures may need extensive debugging to locate the faults.


The mutators (methods to perform mutations) implemented in our approach
can be divided into three classes (i) simple replacements within a fixed
set of symbols, (ii) replacements of symbols found in the function body,
(iii) deletion of lines. The sets of symbols used for replacement are given
in \Cref{tab: Mutators}. These mutators are \textit{Assignment}, 
\textit{BooleanAssignment}, \textit{Comparison}, \textit{Arithmetic},
\textit{IncrementDecrement}, \textit{BooleanArithmetic}, and \textit{Logical}.
Their purpose is to induce execution to diverge into paths where
the current program state is invalid, e.g. by perturbing conditionals,
and loop increments. We use the Python Tree-sitter\footnote{\url{https://github.com/tree-sitter/tree-sitter}} library to obtain the abstract syntax tree, identify nodes, and apply precise code mutations.

\subsection{Mutation Execution}

We utilize the test suites to execute the mutated source code. Therefore, we select the source code targets for mutation based on test case coverage information.
For the selected mutation targets, we apply the set of mutators as described. We present an overview of our mutation process in \Cref{fig:MutationProcess}. Then we build the source code. We discard all build failures. We then execute the code through test frameworks.
If the mutation leads to a program receiving a signal like SIGSEGV, resulting in the abrupt termination of the program as illustrated in step 3 of \Cref{fig: promptExample}, due to an invalid memory operation, we consider such a scenario as a crash. Only mutations that lead to crashes, defined as the executable receiving a signal from the operating system, are considered for evaluation.  Once the mutated executable crashes, a stack trace is extracted with the debugger and recorded into our dataset. Finally, the mutation is rolled back in the source code, and we perform a new mutation. 

For mutation, we primarily focus on SAP HANA~\cite{farber2012sap} an enterprise-level ERP database system implemented in C++. Additionally, we choose two open-source database projects: SQLite~\cite{sqlite_hipp2000} and DuckDB~\cite{raasveldt2019duckdb}. DuckDB, developed in C++, is optimized for in-memory analytical queries. SQLite, written in C, is a lightweight, serverless DBMS commonly used in embedded systems. 

We enable debug symbols to compile the code for SQLite and DuckDB and run their respective test suites under GNU GDB to collect traces for crashes. HANA has its own internal unwinding framework that provides stack traces.  We ignore the mutations that lead to i) an ordinary completion of the shell executable or ii) execution times that exceed a predetermined threshold. We determine the threshold duration for each codebase separately since successful build and test runtimes differ by project. We record the time for a successful build under the specific settings used for each project. This recorded time then is set as the build threshold. As some mutations can cause longer runtimes exceeding the threshold times, we terminate such builds. In the final dataset, the stack traces resulting from the crash are mapped to the mutated method.

\subsection{Types of Crashes}
Some crashes occur close to the code being executed, making the faulty function identifiable within the stack trace and relatively easy to localize. However, even in these cases, the faulty function may appear at any depth in the stack trace, adding complexity to localization. More challenging are cases where the faulty method doesn’t appear in the stack trace at all, making localization significantly more difficult.
Based on these aspects of the complexity of fault localization, we categorize the resulting crashes into two types. i) \textit{Local crashes}: Crashes where the mutated function name is found within the resulting stack trace.  ii) \textit{Non-local crashes}: Crashes where the mutated function name is not included in the stack trace.

\subsection{Preprocessing Stack Traces}
We choose to fine-tune the open-source LLMs. With the open-source LLMs, we are constrained with respect to the available context length, and preprocessing allows us to retain the most important parts of the stack traces. Step 4 of \Cref{fig: promptExample} illustrates an example of the preprocessed SQLite stack trace. The stack trace shows the active function calls during a system crash and the exact location of the crash as shown in \Cref{fig: promptExample}.
Stack traces consist of static components like method names, class names, and line numbers, which are useful for identifying faults, and dynamic components like memory addresses, thread identifiers, timestamps, and register values that vary with each execution. We exclude the dynamic components during preprocessing because they do not significantly contribute to fault localization. The frame order can differ based on the collection method. The GNU GDB returns a different order than that of HANA's method. We ensure the most recent call is at the beginning of the stack trace. In the case of HANA, the crash dumps contain a large and dense amount of information on the database's state, runtime information, etc. We extract parts that contain the stack traces of the crash. 

After preprocessing, SQLite and HANA retain the full depth of stack traces, while in DuckDB, truncation mostly eliminates frames pertaining to the test execution framework.
The preprocessing steps we apply may cause stack traces resulting from different mutation locations to appear identical, thereby reducing the effectiveness of the fine-tuning process. Therefore, we filter the dataset to retain distinct stack traces to function mappings.  In the SQLite and DuckDB datasets, the average occurrence of such cases is 1, whereas in the HANA dataset, the average occurrence is 2.37. In HANA, we deduplicate the dataset based on the preprocessed stack traces.

\section{Prediction Task}
\label{s: prediction_task}

\floatstyle{plain}
\newfloat{listing}{ht}{lop}
\floatname{listing}{Listing}
\lstset{basicstyle=\ttfamily\tiny,breaklines=true}

\begin{listing}[htbp]
\caption{Example Prompt for SQLite and Llama Instruct.  }
\begin{tcolorbox}[colback=white, colframe=black, fonttitle=\bfseries, boxrule=0.3pt,  title=Prompt, colbacktitle=white!60!black, halign=flush left,
left=2pt, right=2pt, top=2pt, bottom=2pt,]

ROLE: You are a software engineering assistant who can analyze stack traces from a crash and assist a C developer in localizing the fault to a method in the stack trace.\\
TASK: Given a stack trace, identify the function name that most likely caused the crash. The faulty method is not necessarily closer to the last frame in the stack trace.\\
INPUT: Preprocessed stack traces that resulted from a crash in the open-source Database project SQLite, which has been written in C programming language.\\

\text{[CRASH STACK]:}\\
\text{\#3   in main .. at src/tclsqlite.c:4062}\\
\text{\#2   in sqlite3TestInit  ..}\\ 
\text{..}\\
\text{fs\_register() at src/test\_onefile.c:824} \\
\text{Program received signal SIGSEGV, Segmentation fault.}\\ 
OUTPUT: Strictly provide only the function name without any additional information in the format:\\
<function\_name>
\end{tcolorbox}

\begin{tcolorbox}[colback=white, colframe=black, fonttitle=\bfseries, boxrule=0.3pt, title= Response, colbacktitle=white!60!black, halign=flush left,
left=2pt, right=2pt, top=2pt, bottom=2pt,]

\text{<sqlite3\_malloc>}

\end{tcolorbox}
\label{SamplePrompt}
\end{listing}

LLMs are trained on publicly available code and related data. They have been used for fault localization and understanding code and failures through prompting techniques. Code-based LLMs are not as effective at comprehending instructions in a prompt as their instruct-tuned versions. Code-based LLMs often generate tokens unrelated to the expected output when not explicitly trained for a particular task as ours. Therefore, predictions can be made using prompting techniques with general-purpose language models like GPT4-o. Alternatively, we can expect an accurate inference from fine-tuned LLMs.

\subsection{Prompting LLMs}
 We use zero-shot prompting to query faulty methods from instruction-capable LLMs. The general structure of our prompts has a role definition, task description, input description,  and output structure constraints. Initially, we define a role for the LLM as an 'assistant' to the developer who is capable of analyzing stack traces of a software crash to identify the faulty method that may have most likely caused the crash. We provide information on the programming language and name of the database project from which the stack trace has been collected. We describe the input information as preprocessed stack traces collected from a crash. We define the output structure and instruct the LLM to strictly adhere to producing a function name without any additional explanation or sentences. \Cref{SamplePrompt} is an example prompt and response for an SQLite crash caused by mutating \textit{malloc} function.

\subsection{Inference with Finetuned LLMs}
Our fine-tuning task involves a text prediction task, where the LLM predicts the faulty function name with a file path given a preprocessed stack trace.
While fine-tuning an LLM we train the target as a file name with a function name that can comprise multiple tokens. Various methods control token generation, such as setting a maximum token limit or using an end-of-generation token selected during fine-tuning.  We set the end-of-generation token to \textsc{'<|endoftext|>'}. For the prediction phase, we generate the tokens where the model greedily chooses the next most probable token as completion until it encounters an end-of-generation token.

\section{Experiments and Evaluation}

\label{s: expt_and_eval}

In this section, we discuss our experiment setup and dataset and define the evaluation metric for each experiment. 
\subsection{Dataset}


\subsubsection{Synthethic Crash Dataset}
The dataset of synthetic crashes in our experiments is presented in \Cref{tab:DataCounts}. The types of mutations applicable vary depending on the programming language the project is implemented in. Therefore, we implement custom mutators that are applicable to specific libraries in HANA. The types of mutations across SQLite and DuckDB remain the same. We do not balance the synthetic dataset w.r.t the function names or the mutation types. Although the process of mutating source code may be influenced by the size of the target functions in terms of lines of code, our resulting dataset includes a large number of unique target functions and shows no significant bias toward specific targets. We deduplicate the dataset before splitting it into training and validation sets, ensuring no data leakage. The data is randomly split into a \num{90}\% training set for fine-tuning the model and the remaining \num{10}\% as a validation set. We use the same validation dataset across all our experiments pertaining to synthetic crashes.
We make our synthetic crash datasets for SQLite and DuckDB available here\footnote{\url{https://doi.org/10.5281/zenodo.14680837}}.


\subsubsection{Authentic Crash Dataset}

Over several years, we have collected data on crashes in HANA. For these crashes, fixes are often complex. Fixes may involve changes to multiple files and functions unrelated to the root cause of the failure. Additionally, changes to code over a decade have resulted in methods that were changed as a part of a fix but no longer exist. To ensure our approach has even a chance of predicting the correct locations, we significantly trim the authentic crashes to retain only crashes where fixes entail changes to a single file but multiple methods, and one of the fixed methods is present in the stack trace. Consequently, we have 175 authentic local crashes spanning five years.

\begin{table}[htbp]
\caption{\#C: count of crashes \#T: unique targets - file path with function names, \%L: percentage of local crashes in data.}
\resizebox{\columnwidth}{!}{
\begin{tabular}{@{}lrrlrrlrrl@{}}
\toprule
 & \multicolumn{3}{c}{Training} & \multicolumn{3}{c}{Validation} & \multicolumn{3}{c}{Authentic} \\ \cmidrule(l){2-10} 
Project & \multicolumn{1}{l}{\#C} & \multicolumn{1}{l}{\#T} & \%L & \multicolumn{1}{l}{\#C} & \multicolumn{1}{l}{\#T} & \%L & \multicolumn{1}{l}{\#C} & \multicolumn{1}{l}{\#T} & \%L \\ \midrule
HANA & 64369 & 8609 & 33.1 & 7152 & 3123 & 32.4 & 175 & 82 & 100 \\
SQLite & 13299 & 1183 & 47 & 1477 & 569 & 46 & - & - & - \\
DuckDB & 1635 & 345 & 55.8 & 181 & 120 & 57 & - & - & - \\ \bottomrule
\end{tabular}
}
\label{tab:DataCounts}
\end{table}



\subsection{Prediction Task}
The task is to predict the faulty method causing the crash.
\subsubsection{On Synthetic Crashes}
\label{sss: SynCrashPred}
In fine-tuned models, the prediction must accurately match the file name and the method name. We introduce this output structure to pre-trained LLMs through fine-tuning. To evaluate the impact of fine-tuning, we restrict inference to a single prediction, as expanding to top-\textit{k} sampling may only improve performance. In non-fine-tuned LLMs, the models must predict only the function name.  

\subsubsection{On Authentic Crashes}
Authentic crashes have complex fixes, which consist of modifications to $n$ different methods. If at least one of the  $n$ methods is predicted as a fault, we consider it accurately localized.

\subsection{Models and Setup}

\subsubsection{Non-fine-tuned LLMs}
\label{ss: InstructModels}

 Non-fine-tuned LLMs may not recognize the necessary file names or understand the required output structure. Additionally, non-instruct versions of LLMs typically are not designed for comprehending instructions in prompts, where we define a task and desired output format. Therefore, we choose to use the instruct variations of LLMs for our non-fine-tuned LLM experiments - specifically, Mistral-7b-Instruct (Mistral-I)~\cite{jiang2024mixtral}, LLAMA-70B-Instruct (Llama-I)~\cite{touvron2023llama}, and  GPT-4o\footnote{\url{https://openai.com/index/hello-gpt-4o/}}, one of the best LLMs across several tasks. Mistral-I outperforms Llama-I across several benchmarks\footnote{\url{https://mistral.ai/news/mixtral-of-experts/}}, Llama-I is an instruct model well suited for coding tasks in the Llama family\footnote{\url{https://deepinfra.com/meta-llama/Llama-2-70b-chat-hf}}.

Non-fine-tuned models trained on public datasets may possess an inherent understanding of stack traces and may be inclined to pick a function name from within the stack trace. We noticed this pattern with our manual experiments with prompting. Therefore, we modify our prediction task to identify the faulty function name and also limit our dataset to the \textit{local crashes} where the faulty method is present in the stack trace. We consider a target accurately predicted when the LLM picks the right function name from the input stack trace. Every LLM may comprehend instructions differently and require separate prompt optimization across each codebase to obtain the desired output. Therefore, we manually optimize the prompt for each LLM and project by examining 10 sample cases. We set the temperature to 0 across all these models.

\subsubsection{Fine-tuned LLMs}
We conduct experiments with three open source LLMs: Mistral-7B~\cite{jiang2023mistral}, Santacoder-1B~\cite{allal2023Santacoder}, and CodeLlama-7B~\cite{roziere2023code}.  Santacoder-1B has not been trained on C and C++ code or failures. However, CodeLlama and Mistral may have been trained with C and C++ code. Santacoder-1B and CodeLlama-7B are particularly suited for code-related tasks, and Mistral-7B is capable of comprehending instructions and performs suitably well on code-related tasks.  We fine-tune CodeLlama-7B and Mistral-7B with LoRA~\cite{hu2021lora} adapters which results in training only about 1\% of the full model parameters during fine-tuning. We limit the input to a maximum of 1024 tokens. This token limit is sufficient to fit the required tokens from pre-processed stack traces across all our models, ensuring uniform data across experiments for model comparison.

 Baselines: i) Innerframe: A method that chooses the \textit{'innermost frame'} in the stack trace. In a sequence of calls, it is likely the most relevant fault location. ii) fasttext with cosine similarity: We train a \textit{fasttext}~\cite{bojanowski2017enriching} embedding model and vectorize the pre-processed stack traces to predict function name using 1-nearest-neighbor(\textit{1-NN}) classifier using cosine similarity. iii) Non-fine-tuned instruction capable LLMs mentioned in \Cref{ss: InstructModels}.

\subsection{Metrics}
\subsubsection{Perfect Match / Accuracy}
We present the \textit{Accuracy} of the models as our evaluation metric. For fine-tuned LLMs, we consider a target accurately predicted when the predicted tokens, excluding the end-of-generation, perfectly match the target tokens. For non-fine-tuned  LLMs, if the model picks the right function from within the stack trace, we consider it accurately predicted.

\subsubsection{Average Precision}
\label{ss: AvgPrecision}

We choose average precision as a metric to assess an LLM for partially identifying the fault location. We consider that correctly locating the right component, class, or file can be helpful in pinpointing faulty code. Therefore, we evaluate the similarity between the predicted and target strings.  To determine the precision, we convert both the prediction and target strings into individual terms. For example, in C++ target string \texttt{'X/Y/Z.cpp A::B::C'} becomes [X, Y, Z, A, B, C]. We disregard the order of terms and compute the percentage of terms in the prediction that match terms in the target. This is similar to calculating the BLEU-1 score~\cite{papineni2002bleu} without the brevity penalty. The average precision of the terms produced by the LLM across all instances is calculated as shown in \Cref{eq: avg_precision_new}.

Average Precision (AP), is calculated as:


\begin{equation}
\text{AP} = \frac{1}{N} \sum_{i=1}^{N} \frac{\text{$|\text{terms}(P_i) \cap\ \text{terms}(T_i)|$  }}{\text{ $|\text{terms }(P_i)|$}}
\label{eq: avg_precision_new}
\end{equation}

where $P_i$ and $T_i$ are the Prediction and Target of the $i$th instance, respectively. $N$  is the total number of predictions.





\section{Discussion}
\label{s: discussion}
In this section, we discuss the results of fault localization across fine-tuning LLMs and prompting instruct capable LLMs.




\subsection{Fault Localization with Fine-tuned LLMs}

\begin{table*}[ht]
\caption{Accuracy of predicting file and faulty function name on Validation Data - Baseline and fine-tuned LLM. \\(Auth: Authentic crashes, Syn: Synthetic crashes, InnerM: InnerMost Baseline)}
\begin{tabular}{@{}lrrrrrrrrrrrrrrr@{}}
\toprule
\multicolumn{1}{c}{} & \multicolumn{1}{l}{} & \multicolumn{1}{l}{} & \multicolumn{13}{c}{Accuracy} \\ \midrule
Models & \multicolumn{1}{l}{} & \multicolumn{1}{l|}{} & \multicolumn{3}{c|}{Santacoder-1B} & \multicolumn{3}{c|}{Mistral-7B} & \multicolumn{3}{c|}{CodeLlama-7B} & \multicolumn{3}{c|}{Fasttext Cosine} & \multicolumn{1}{l}{InnerM} \\ \midrule
Project & \multicolumn{1}{l}{Count} & \multicolumn{1}{l|}{\begin{tabular}[c]{@{}l@{}}Max\\ Depth\end{tabular}} & \multicolumn{1}{l}{Local} & \multicolumn{1}{l}{\begin{tabular}[c]{@{}l@{}}Non\\ Local\end{tabular}} & \multicolumn{1}{l|}{Total} & \multicolumn{1}{l}{Local} & \multicolumn{1}{l}{\begin{tabular}[c]{@{}l@{}}Non\\ Local\end{tabular}} & \multicolumn{1}{l|}{Total} & \multicolumn{1}{l}{Local} & \multicolumn{1}{l}{\begin{tabular}[c]{@{}l@{}}Non\\ Local\end{tabular}} & \multicolumn{1}{l|}{Total} & \multicolumn{1}{l}{Local} & \multicolumn{1}{l}{\begin{tabular}[c]{@{}l@{}}Non\\ Local\end{tabular}} & \multicolumn{1}{l|}{Total} & \multicolumn{1}{c}{Total} \\ \midrule
SQLite-Syn & 1082 & \multicolumn{1}{r|}{13} & 0.758 & 0.375 & \multicolumn{1}{r|}{0.557} & 0.832 & 0.422 & \multicolumn{1}{r|}{\textbf{0.634}} & 0.744 & 0.271 & \multicolumn{1}{r|}{0.515} & 0.558 & 0.434 & \multicolumn{1}{r|}{0.558} & 0.192 \\
DuckDB-Syn & 181 & \multicolumn{1}{r|}{7} & 0.597 & 0.261 & \multicolumn{1}{r|}{0.453} & 0.854 & 0.587 & \multicolumn{1}{r|}{\textbf{0.741}} & 0.725 & 0.283 & \multicolumn{1}{r|}{0.537} & 0.709 & 0.565 & \multicolumn{1}{r|}{0.629} & 0.204 \\ \midrule
HANA-Syn & 7152 & \multicolumn{1}{r|}{59} & 0.795 & 0.489 & \multicolumn{1}{r|}{0.588} & 0.884 & 0.545 & \multicolumn{1}{r|}{0.655} & 0.879 & 0.568 & \multicolumn{1}{r|}{\textbf{0.669}} & 0.133 & 0.093 & \multicolumn{1}{r|}{0.106} & 0.126 \\
HANA-Auth & 175 & \multicolumn{1}{r|}{104} & 0.354 & - & \multicolumn{1}{r|}{\textbf{0.354}} & 0.297 & - & \multicolumn{1}{r|}{{0.297}} & 0.297 & - & \multicolumn{1}{r|}{0.297} & 0.00 & - & \multicolumn{1}{r|}{0.00} & 0.185 \\ \bottomrule
\end{tabular}

\label{tab: Accuracy}
\end{table*}

\subsubsection{Performance on Synthetic Crashes}
\Cref{tab: Accuracy} presents an overview of the overall performance for all the fine-tuned LLMs and the baselines \textit{fasttext} and \textit{Innermost} across all the codebases. While fine-tuning Santacoder-1B on HANA crashes, we observed evolving validation set accuracies as seen in \Cref{fig: TestVsEpochsAcc}. Continued training improved the accuracy for non-local crashes, but these remain more challenging compared to local crashes. This trend is consistent across all our projects and LLM combinations. As shown in \Cref{tab: Accuracy}, non-local crashes consistently have lower accuracies than local crashes. 


\begin{figure}

\includegraphics[scale=0.66]{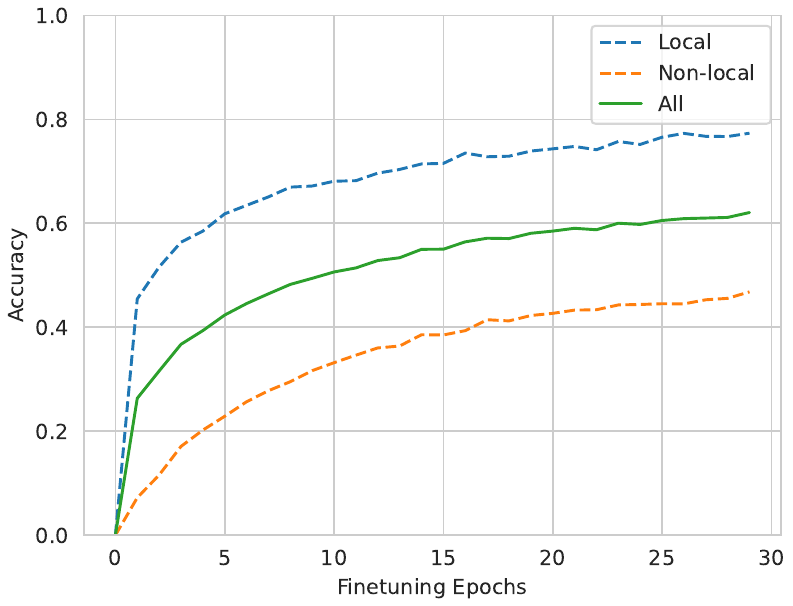} 
\caption{HANA Validation Accuracy on Santacoder-1B.}
\label{fig: TestVsEpochsAcc}
\end{figure}

In our experiments, we observed no correlation between the depth at which the mutated function name is present and the LLMs' effectiveness in localizing faults within the validation data. To determine if the LLM relies heavily on learning associations with function names present in the stack trace, we analyzed the misclassifications to identify if the predicted function names were included in the prompt. However, this was not the case. All LLMs tended to make predictions without a particular bias towards function names from the stack trace, across both local and non-local crash misclassifications. We also found no correlation between the length of stack traces and the accuracy of fault localization across all our experiments with fine-tuned models. We observed that LLMs generated accurate file paths for some of the misclassified instances. As shown in \Cref{fig: promptExample}, LLMs are required to generate both the file path and function names. Therefore, when considering file-level localization, we observed an average increase in overall localization accuracy by 5\% in DuckB (to 62\%), 34\% in SQLite (to 89\%), and 10\% in HANA (to 76\%) across all LLMs. This improvement can be due to the ease of a coarse-grained task of predicting file names.



The LLMs we fine-tuned are of different sizes, pre-trained on different data, and possess different inherent capabilities. For instance, Santacoder-1B is not inherently aware of C and C++ style code and stack traces. In our experiments with HANA and SQLite on fine-tuned Santacoder-1B, a model with just 1 billion parameters, we observe that it performs on par with larger models like CodeLlama-7B and Mistral-7B. We can partly attribute the comparatively better performance across  HANA and SQLite to the relatively larger dataset we were able to gather for fine-tuning. 

\subsubsection{Perfomance of fasttext}The fasttext baseline identifies the root cause functions on par with the LLMs across SQLite and DuckDB synthetic crashes. However, this approach has a low accuracy of 10\% on HANA synthetic crashes and is unable to localize any of the authentic HANA crashes despite having a complete overlap of the target functions of the authentic crashes with the training set. \textit{fasttext} primarily treats text as a bag-of-words, disregarding its sequential structure. Hence it doesn't effectively capture long-range dependencies. LLMs, being transformer-based architectures, excel at learning representations for long sequences, effectively preserving the structural nuances of stack traces~\cite{vaswani2017attention}. As shown in \Cref{tab: Accuracy}, the performance disparity between the LLMs and fasttext on HANA highlights the influence of stack trace structure on HANA compared to SQLite and DuckDB.






\begin{table}[]
\centering
\caption{Correct predictions for targets that were not present in the training set.  UT: Unseen Targets during fine-tuning. }
\begin{tabular}{@{}lrrrr@{}}
\toprule
 & \multicolumn{1}{l}{} & \multicolumn{3}{c}{\#Correct Predictions on UT} \\ \cmidrule(l){3-5} 
Project & \#UT & Santacoder-1B & Mistral-7B & \multicolumn{1}{l}{CodeLlama-7B} \\ \midrule
SQLite-Syn & 25 & 4 & 2 & 2 \\
DuckDB-Syn & 14 & 1 & 4 & 0 \\ \bottomrule
\end{tabular}
\label{tab: UknClass}
\end{table}


\subsubsection{Unseen Targets} Random sampling without stratification for targets resulted in validation sets for DuckDB and SQLite with stack traces of certain mutated functions and, consequently, the function names are not present in the training set.  As shown in \Cref{tab: UknClass}, LLMs predicted some of such unseen targets correctly. Our approach benefits from the generative aspect of LLMs being capable of abstracting inherent patterns and interpolating unseen but correct locations, as shown in the case of SQLite and DuckDB. \\

\subsubsection{Performance on Authentic Crashes} Even with a limited set of authentic data, we show in \Cref{tab: Accuracy} that LLMs fine-tuned on synthetic crash data can localize authentic bugs with an accuracy of 35.4\%. Despite relatively weaker performance on synthetic crashes, Santacoder-1B outperforms CodeLlama-7B and Mistral-7B on the authentic crash dataset. This could be an indication of larger models overfitting our synthetic crash dataset. Based on average precision in \Cref{tab: AvgPrecision}, fine-tuned LLMs produced more useful information across all experiments in comparison to non-fine-tuned LLMs.

\subsection{Analysis across Mutation Operators}

 We analyze the crashes caused by different mutators and evaluate the accuracy of localizing them. Our data shows that certain mutators cause more crashes and complex failures than others.  In the HANA dataset, the distribution of crashes from different mutation operators is consistent across training and test splits as shown in \Cref{fig: TrTestMutDistAcc}. The top 3 mutators result in 86.8\% of all crashes and 85\% of wrongly localized crashes in the HANA dataset. 78\% non-local crashes of comparison mutators are misclassified in HANA. Research indicates that Drop Line mutants often reflect real faults~\cite{just2014mutants}.  Drop Line operator constitutes 30\% of all crashes in the HANA dataset. 72\% of the Drop Line mutation crashes are non-local crashes. In HANA, 52\% of the wrong predictions belong to non-local crashes of the Drop Line operator. We found that mutations to lines of code containing assignments, method calls, and pointer operations resulted in more non-local crashes than other types of code. Similarly, in both SQLite and DuckDB datasets, Drop Line and comparison mutators lead to the most non-local crashes. Across SQLite and DuckDB datasets, non-local crashes resulting from Drop Line have an accuracy of 0\%, whereas local crashes are localized with an accuracy of 100\%. Across all the other mutators, we do not see a distinct pattern of misclassification in our datasets. DropLine and comparison operators induce complex failures and stack traces alone may be insufficient to localize such faults.


\begin{figure*}[ht]
  
    \begin{subfigure}[b]{0.68\columnwidth}
        \includegraphics[width=\textwidth]{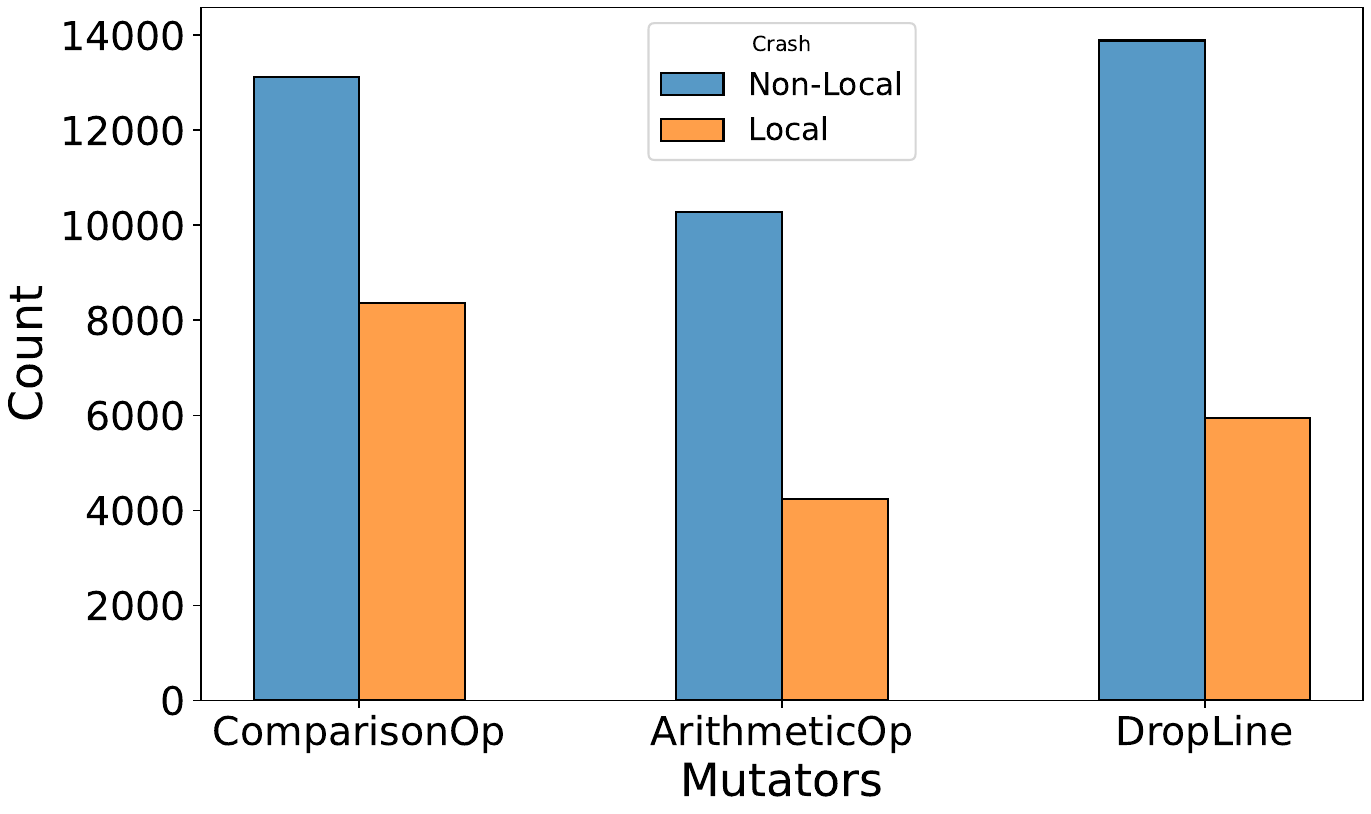}
        \caption{Training data }
        \label{fig: TrainDataHANA}
    \end{subfigure}
    \begin{subfigure}[b]{0.68\columnwidth}
        \includegraphics[width=\textwidth]{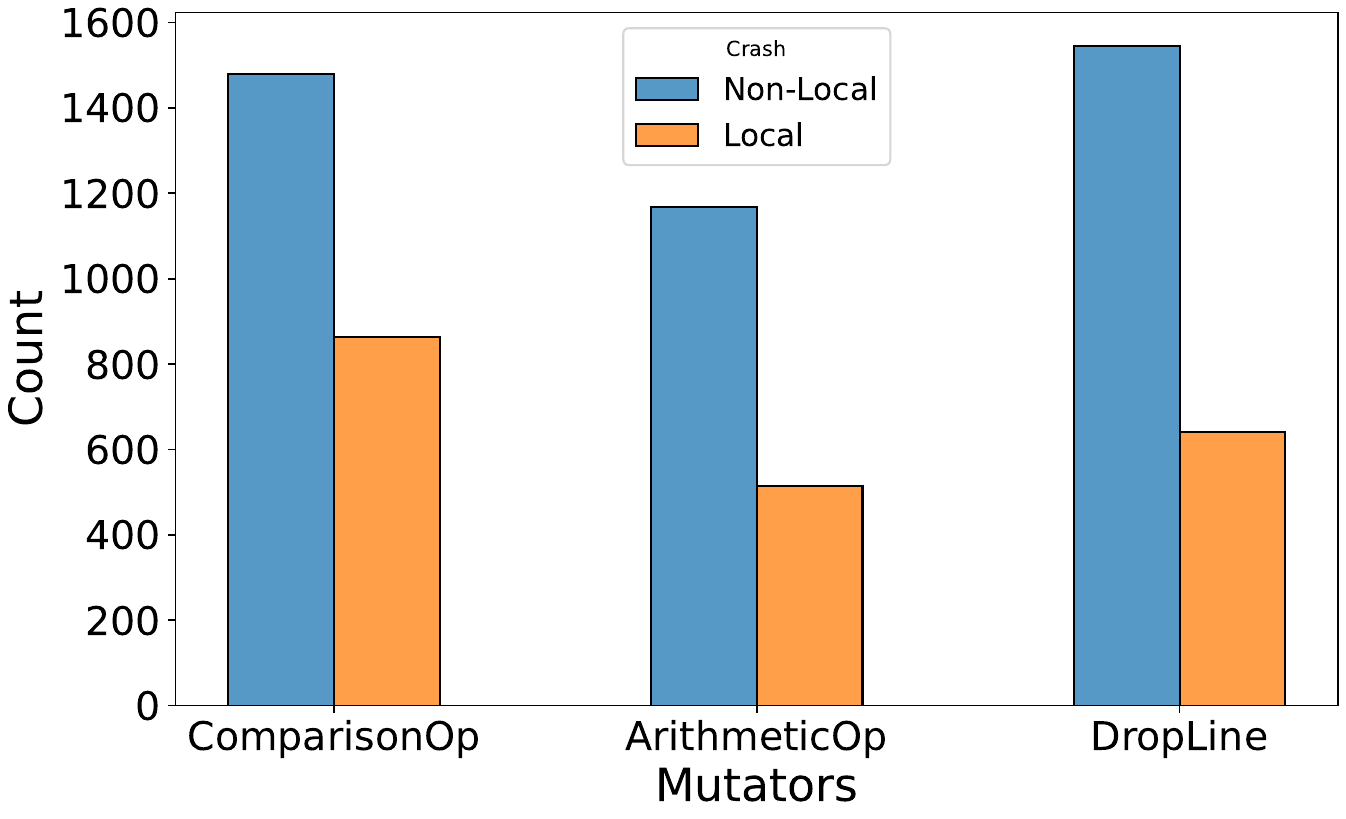}
        \caption{Test data }
        \label{fig: TestDataHana}
    \end{subfigure}
    \begin{subfigure}[b]{0.67\columnwidth}
        \includegraphics[width=\textwidth]{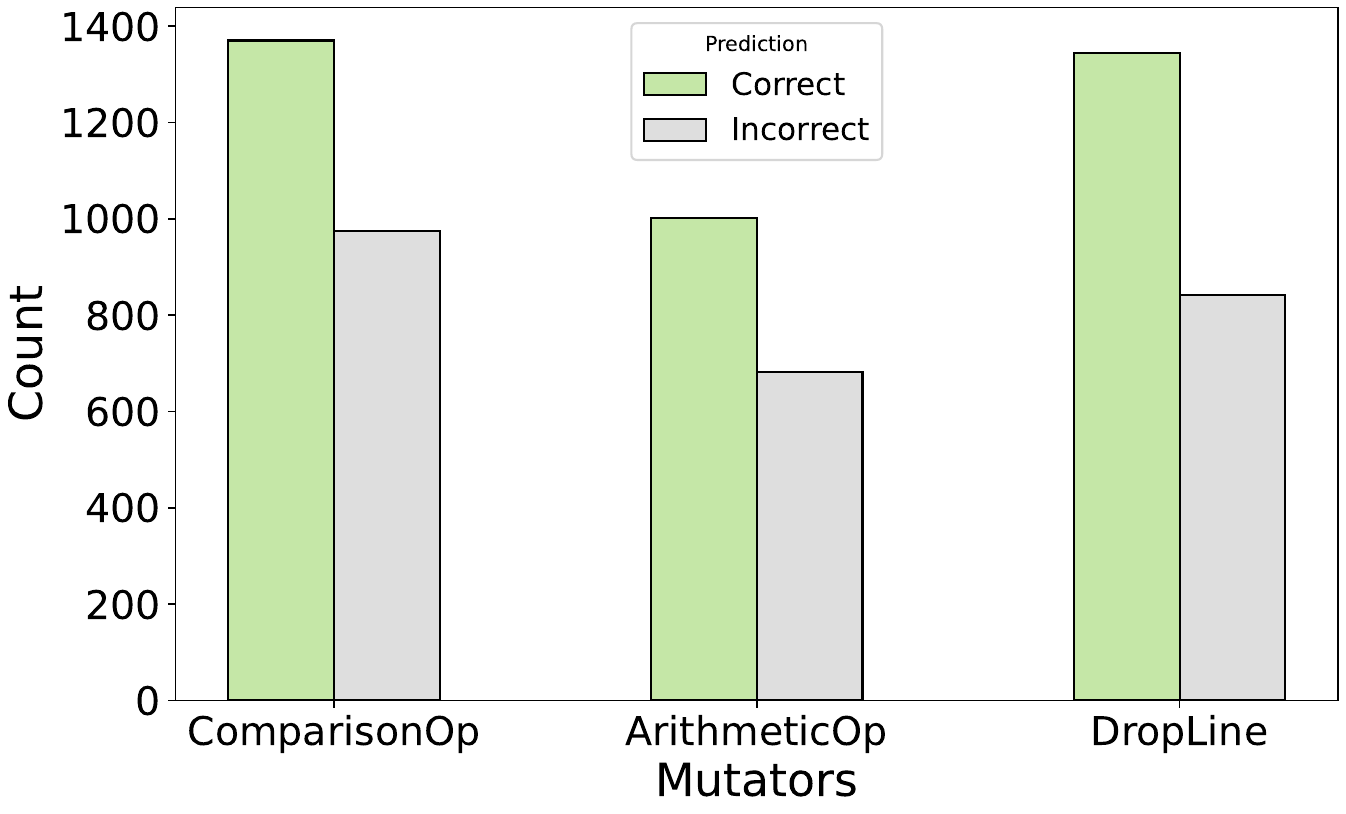}
        \caption{Accuracy with Santacoder-1B}
        \label{fig: AccMutators}
    \end{subfigure}
    \caption{Top 3 Mutator distribution in Synthetic HANA Dataset.}
    \label{fig: TrTestMutDistAcc}
\end{figure*}

\subsection{Fault Localization with Prompting Non-fine-tuned LLMs}
 We evaluate the performance of the Instruct LLMs for localizing faults i) Based only on stack traces, ii) Based on stack traces augmented with function implementations, and iii) Based on obfuscated stack traces. We also evaluate them for the usefulness of their output with an average precision.

\begin{table*}[htbp]
\centering
\caption{Accuracy of Non-fine-tuned and Fine-tuned LLMs on local crashes. Only Local crashes are used for this experiment. 
(Auth: Authentic crashes, Syn: Synthetic crashes, FD: Prompts augmented with function source code)}
\begin{tabular}{@{}lrrrrrrr@{}}
\toprule
\multicolumn{1}{c}{} & \multicolumn{1}{l}{} & \multicolumn{3}{c|}{Accuracy Non-Fine-tuned} & \multicolumn{3}{c}{Accuracy Fine-tuned} \\  \cmidrule(l){3-8} 
Models & \multicolumn{1}{l|}{} & GPT-4o & Mistral-I & \multicolumn{1}{r|}{Llama-I} & \multicolumn{1}{l}{Santacoder-1B} & \multicolumn{1}{l}{Mistral-7B} & \multicolumn{1}{l}{CodeLlama-7B} \\ \midrule
Project & \multicolumn{1}{l|}{Count} & \multicolumn{3}{c|}{Local} & \multicolumn{3}{c}{Local} \\ \midrule
SQLite-Syn & \multicolumn{1}{r|}{680} & \textbf{0.567} & 0.357 & \multicolumn{1}{r|}{0.496} & 0.758 & \textbf{0.832} & 0.744 \\
DuckDB-Syn & \multicolumn{1}{r|}{103} & 0.0 & \textbf{0.161} & \multicolumn{1}{r|}{0.0} & 0.597 & \textbf{0.854} & 0.725 \\ \midrule
HANA-Syn & \multicolumn{1}{r|}{2317} & 0.212 & 0.241 & \multicolumn{1}{r|}{\textbf{0.422}} & 0.795 & \textbf{0.884} & 0.879 \\
HANA-Auth & \multicolumn{1}{r|}{175} & 0.091 & 0.12 & \multicolumn{1}{r|}{\textbf{0.229}} & \textbf{0.354} & 0.297 & 0.297 \\
HANA-Syn-FD & \multicolumn{1}{r|}{1469} & 0.094 & - & \multicolumn{1}{r|}{-} & - & - & - \\ \bottomrule
\end{tabular}

\label{tab: unfinetunedLLM}
\end{table*}

\subsubsection{Prompts with only Stack Traces}
We prompt the LLMs as shown in \Cref{SamplePrompt} with preprocessed stack traces. Llama-I outperformed Mistral-I and GPT-4o on the HANA dataset of local crashes. Across DuckDB Mistral has 12\% while the other LLMs have 0\%. Despite considering only local crashes, the accuracy across all the LLMs and projects is consistently poor as shown in \Cref{tab: unfinetunedLLM} in comparison to their fine-tuned counterparts.  \Cref{tab: unfinetunedLLMStackDepth} shows the average depths at which the LLMs selected the faulty method. If the LLM chose the first frame, the depth would be considered as 0, and if it picked the last, then the depth would be 1. We exclude instances where the LLMs returned unparseable outputs for average depth calculation. The stack traces include functions from the test framework and exception-handling frameworks that may have caught the failures. In SAP HANA, such frames tend to be at the beginning and, at times, also at the end of the stack traces. In DuckDB and SQLite, such frames tend to be at the beginning.  In most cases, LLMs bypassed such tests and stack unwinding related frames. We analyzed how the length of stack traces affects accuracy and found no correlation between them.

\begin{table}[htbp]
\centering
\caption{Average depth at which Non-fine-tuned LLMs predicted the faulty function from the input stack traces.\\(Syn: Synthetic crashes)} 
\begin{tabular}{@{}lrrrrrr@{}}
\toprule
\multicolumn{1}{c}{} & \multicolumn{6}{c}{Avg. Stack Depth of Prediction} \\ \midrule
\multicolumn{1}{l}{Models} & \multicolumn{2}{r|}{GPT-4o} & \multicolumn{2}{r|}{Mistral-I} & \multicolumn{2}{r}{Llama-I} \\ \midrule
\multicolumn{1}{l|}{Project} & Local & \multicolumn{1}{r|}{\begin{tabular}[c]{@{}r@{}}Non\\ Local\end{tabular}} & Local & \multicolumn{1}{r|}{\begin{tabular}[c]{@{}r@{}}Non\\ Local\end{tabular}} & Local & \begin{tabular}[c]{@{}r@{}}Non\\ Local\end{tabular} \\ \midrule
\multicolumn{1}{l|}{SQLite-Syn} & 0.93 & \multicolumn{1}{r|}{0.91} & 0.15 & \multicolumn{1}{r|}{0.01} & 0.27 & 0.28 \\
\multicolumn{1}{l|}{DuckDB-Syn} & 0.93 & \multicolumn{1}{r|}{0.93} & 0.27 & \multicolumn{1}{r|}{0.21} & 0.27 & 0.23 \\
\multicolumn{1}{l|}{HANA-Syn} & 0.49 & \multicolumn{1}{r|}{0.50} & 0.47 & \multicolumn{1}{r|}{0.51} & 0.37 & 0.42 \\ \bottomrule
\end{tabular}
\label{tab: unfinetunedLLMStackDepth}
\end{table}

\begin{table*}[htbp]
\centering
\caption{Average precision of prediction measuring the usefulness of the predictions across fine-tuned and non-fine-tuned LLMs. (Auth: Authentic crashes, Syn: Synthetic crashes, FD: Prompts augmented with function source code)}
\begin{tabular}{@{}llrrrrrr|rrrrrr@{}}
\toprule
\multicolumn{1}{c}{} &  & \multicolumn{6}{c|}{Avg. Precision Non Fine-tuned \%} & \multicolumn{6}{c}{Avg. Precision Fine-tuned \%} \\ \cmidrule(l){3-14} 
\multicolumn{1}{l}{Models} &  & \multicolumn{2}{c|}{GPT-4o} & \multicolumn{2}{c|}{Llama-I} & \multicolumn{2}{c|}{Mistral-I} & \multicolumn{2}{c|}{Santacoder-1B} & \multicolumn{2}{c|}{CodeLlama-7B} & \multicolumn{2}{c}{Mistral-7B} \\ \midrule
\multicolumn{1}{l|}{Project} & \multicolumn{1}{l|}{Count} & Local & \multicolumn{1}{r|}{\begin{tabular}[c]{@{}r@{}}Non\\ Local\end{tabular}} & Local & \multicolumn{1}{r|}{\begin{tabular}[c]{@{}r@{}}Non\\ Local\end{tabular}} & Local & \begin{tabular}[c]{@{}r@{}}Non\\ Local\end{tabular} & Local & \multicolumn{1}{r|}{\begin{tabular}[c]{@{}r@{}}Non\\ Local\end{tabular}} & Local & \multicolumn{1}{r|}{\begin{tabular}[c]{@{}r@{}}Non\\ Local\end{tabular}} & Local & \begin{tabular}[c]{@{}r@{}}Non\\ Local\end{tabular} \\ \midrule
\multicolumn{1}{l|}{SQLite-Syn} & \multicolumn{1}{l|}{1082} & \textbf{0.566} & \multicolumn{1}{r|}{0.0} & 0.495 & \multicolumn{1}{r|}{0.0} & 0.351 & 0.0 & 0.894 & \multicolumn{1}{r|}{0.705} & 0.881 & \multicolumn{1}{r|}{0.674} & \textbf{0.936} & \textbf{0.761} \\
\multicolumn{1}{l|}{DuckDB-Syn} & \multicolumn{1}{l|}{108} & \textbf{0.427} & \multicolumn{1}{r|}{0.014} & 0.308 & \multicolumn{1}{r|}{0.014} & 0.278 & \textbf{0.018} & 0.851 & \multicolumn{1}{r|}{0.699} & 0.881 & \multicolumn{1}{r|}{0.711} & \textbf{0.922} & \textbf{0.829} \\ \midrule
\multicolumn{1}{l|}{HANA-Syn} & \multicolumn{1}{l|}{7152} & 0.37 & \multicolumn{1}{r|}{0.42} & \textbf{0.642} & \multicolumn{1}{r|}{\textbf{0.450}} & 0.379 & 0.429 & 0.924 & \multicolumn{1}{r|}{0.759} & 0.940 & \multicolumn{1}{r|}{0.802} & 0.959 & 0.792 \\
\multicolumn{1}{l|}{HANA-Auth} & \multicolumn{1}{l|}{175} & 0.308 & \multicolumn{1}{r|}{-} & \textbf{0.534} & \multicolumn{1}{r|}{-} & 0.302 & - & \textbf{0.640} & \multicolumn{1}{r|}{-} & 0.612 & \multicolumn{1}{r|}{-} & 0.610 & - \\
\multicolumn{1}{l|}{HANA-Syn-FD} & \multicolumn{1}{l|}{1469} & 0.49 & \multicolumn{1}{r|}{0.37} & - & \multicolumn{1}{r|}{-} & - & - &  & \multicolumn{1}{r|}{-} & - & \multicolumn{1}{r|}{-} & - & - \\ \bottomrule
\end{tabular}

\label{tab: AvgPrecision}
\end{table*}

\subsubsection{Prompts Augmented with Function Source-code}
We assessed if adding function implementation code to pre-processed stack traces could improve performance, as SAP HANA is a closed-source system. Function implementations may provide useful comments that enhance the model's understanding. We ignore the function implementations of unit tests and stack unwinding frameworks. We retain 1469 local and 234 non-local crashes where we could reliably extract all important function definitions. Since adding function implementations to the prompt requires longer context lengths than the context length supported by Mistral-I and Llama-I, we only conducted these experiments with GPT-4o. GPT-4o accurately localized only 9.4\% of the faults in local crashes as shown in HANA-Syn-FD \Cref{tab: unfinetunedLLM}. This approach requires accurate source code elements and performs worse than prompts with only stack traces. With the average precision, GPT-4o provided useful information for 720 local crashes as shown in \Cref{tab: AvgPrecision}, which is worse than only using the stack traces. 
Overall, adding function implementations did not improve fault localization.

\subsubsection{Prompts with Stack Trace Obfuscation}

We have seen in our experiments that LLMs understand the basic structure of a stack trace. However, we wanted to investigate if LLMs derive meaning from human-readable elements, specifically method names within the stack trace. To test this, we obfuscated stack traces by using the SHA-256~\cite{pub2012secure} algorithm to hash each line. We informed the LLMs that the stack trace content had been hashed. We tested GPT-4o on 10 SAP HANA local crash examples with these obfuscated stack traces. GPT-4o consistently identified the most frequently repeating line as the culprit method for the crash and provided the same explanation each time. This behavior remained the same even when we separated and hashed each term within the lines individually while keeping the overall line structure. The accuracy of predictions was 0\% across both these experiments. While with unobfuscated prompts the LLM was able to predict the locations correctly. This suggests that LLMs derive information from human-readable and meaningful terms in stack traces.

\subsubsection{Performance on Authentic Crashes}
Llama-I outperformed all other non-fine-tuned models and accurately localized 22.9\% of our authentic crash dataset, while GPT-4o and Mistral-I were at 9.1\% and 12\%, respectively. However, non-fine-tuned models do not match the performance of any of the fine-tuned models across all our experiments as seen in \Cref{tab: unfinetunedLLM} with an accuracy of 35.4\%.

\subsection{Evaluation with Average Precision}
Average Precision, as described in Section \ref{ss: AvgPrecision}, evaluates the LLM's ability to match terms within the target function name. This allows us to measure performance for both local and non-local crashes for non-fine-tuned models.  Even though the accuracy across DuckDB is 0\% for both GPT-4o and Llama-I as presented in \Cref{tab: unfinetunedLLM}, the models were able to correctly predict parts of the fault location, as indicated by their average precision match in \Cref{tab: AvgPrecision}. Fine-tuned LLMs outperform non-fine-tuned LLMs in average precision in all experiments.





\subsection{Limitations}

Our data generation process uses simple mutations, leading to a dataset biased towards less complex failures. Additionally, our authentic dataset of crashes is filtered to include only changes to a single function and function names present in the stack trace, biasing the real-world evaluation towards easier cases. As a result, our evaluation outcomes may be an overestimation of the efficacy of our approach. 


As the source code evolves, the fine-tuned models may become less reliable. Therefore, we need to monitor the accuracy of the predictions over time. If performance degrades below a threshold, we may need to create new data to retrain and further adapt the fine-tuned models.








\section{Conclusion and Future Work}
\label{s: conclusion}

There are failure scenarios where stack traces are the sole information available for localizing faults. We use mutations as a data generation process and fine-tune LLMs to model this cause and effect on program execution. Despite stack traces being incomplete records of failures, we show the viability and generalizability of exploiting them for fault localization by evaluating our approach across multiple LLMs and multiple code bases developed in different programming languages.      

Refining and chaining prompts may serve as an alternative to improving the performance of instruct-capable LLMs. However, even when the information is present in the input, the LLMs are not able to clearly associate it with the source of failure. Our experiments with obfuscation indicate dependency on the inherent meaning of the human-readable information in the stack traces. Additionally, regardless of the prompting strategy, the underlying LLMs never learn the domain-specific knowledge required for our purpose. While fine-tuning involves creating data and training an LLM, it produces more consistent results as it derives from the patterns of mapping between a failure and the fault location learned during the training phase.

By exploring the inner workings of open-source models' attention mechanisms, we can understand how models link tokens in stack traces to predictions. Preprocessing stack traces and performing ablation studies would allow us to identify key elements crucial for fault localization. Incorporating log information and leveraging the multimodal aspects of LLMs can enhance this approach. Additionally, a pretraining step on the entire project code base before fine-tuning can further improve the model's effectiveness. As stated in \Cref{sss: SynCrashPred}, extending our approach to predict top-\textit{k} root cause locations could benefit developers in real-world settings, enhancing the practical utility of our method.
With this research we aim to develop a general approach to handle stack traces across different programming languages, enhancing the adaptability of LLMs. Given their capability in code generation, effective fault localization naturally extends to automatic program repair.

Mutating large codebases is time-consuming and resource-intensive. Only about 10\% of our mutations result in crashes. By fine-tuning an LLM on crashing mutations from our dataset, we can generate context-aware mutators. This approach may eliminate the need to define mutators upfront and could make data generation more efficient.

We currently have made fault localization available to SAP HANA developers, allowing them to input crash stack traces and receive predictions for the faulty method. However, as we are still in the early stages of gathering feedback and optimizing the service, we cannot yet draw conclusions about our solution's performance in day-to-day work.



\balance
\bibliographystyle{IEEEtran}
\bibliography{references}

\end{document}